\documentclass[conference]{IEEEtran}
\IEEEoverridecommandlockouts

\usepackage[T1]{fontenc}
\usepackage{cite}
\usepackage{amsmath,amssymb,amsfonts}
\usepackage{algorithmic}
\usepackage{algorithm}
\usepackage{graphicx}
\usepackage[caption=false,font=footnotesize]{subfig}
\usepackage{textcomp}
\usepackage{pifont}
\usepackage{xspace}
\newcommand{\stepicon}[1]{\ifcase#1\relax\or\ding{182}\or\ding{183}\or\ding{184}\or\ding{185}\or\ding{186}\or\ding{187}\or\ding{188}\or\ding{189}\or\ding{190}\fi}
\newcommand{\BfPara}[1]{{\noindent\bf#1.}\xspace}
\newcommand{\RQ}[1]{\begingroup\setlength{\fboxsep}{1pt}\mbox{\colorbox{gray!20}{\strut\textbf{\textsf{RQ#1}}}}\endgroup}
\newcommand{\RQBox}[2]{\par\vspace{0.5pt}\noindent\begingroup\setlength{\fboxsep}{3pt}\colorbox{rqboxbg}{\hbox{{\color{rqboxaccent}\vrule width 2.5pt}\hspace{3pt}\parbox{\dimexpr\columnwidth-2\fboxsep-5.5pt\relax}{\small\RQ{#1}\hspace{0.35em}#2}}}\endgroup\par\vspace{0.5pt}}
\usepackage{xcolor}
\definecolor{rqboxbg}{RGB}{225,225,250}
\definecolor{rqboxaccent}{RGB}{96,96,180}
\usepackage{booktabs}
\usepackage{array}
\usepackage{url}
\usepackage[breaklinks=true,colorlinks=true,linkcolor=blue,citecolor=blue,urlcolor=blue]{hyperref}
\hypersetup{pdftitle={Beyond Bot Labels: Retrospective Consequence-Grounded Ranking of Attributed Cache Harm in E-Commerce Logs},pdfauthor={Jisoo Kim, Youngwoo Seo, Seonghoon Jeong}}
\usepackage[nameinlink]{cleveref}
\crefformat{section}{#2\S#1#3}
\crefformat{subsection}{#2\S#1#3}
\crefformat{subsubsection}{#2\S#1#3}
\crefmultiformat{section}{#2\S#1#3}{ and~#2\S#1#3}{, #2\S#1#3}{, and~#2\S#1#3}
\crefmultiformat{subsection}{#2\S#1#3}{ and~#2\S#1#3}{, #2\S#1#3}{, and~#2\S#1#3}
\crefmultiformat{subsubsection}{#2\S#1#3}{ and~#2\S#1#3}{, #2\S#1#3}{, and~#2\S#1#3}
\Crefformat{section}{#2\S#1#3}
\Crefformat{subsection}{#2\S#1#3}
\Crefformat{subsubsection}{#2\S#1#3}
\Crefmultiformat{section}{#2\S#1#3}{ and~#2\S#1#3}{, #2\S#1#3}{, and~#2\S#1#3}
\Crefmultiformat{subsection}{#2\S#1#3}{ and~#2\S#1#3}{, #2\S#1#3}{, and~#2\S#1#3}
\Crefmultiformat{subsubsection}{#2\S#1#3}{ and~#2\S#1#3}{, #2\S#1#3}{, and~#2\S#1#3}
\crefname{figure}{Fig.}{Figs.}
\Crefname{figure}{Fig.}{Figs.}
\crefname{table}{Table}{Tables}
\Crefname{table}{Table}{Tables}
\crefname{algorithm}{Algorithm}{Algorithms}
\Crefname{algorithm}{Algorithm}{Algorithms}
\def\BibTeX{{\rm B\kern-.05em{\sc i\kern-.025em b}\kern-.08em
    T\kern-.1667em\lower.7ex\hbox{E}\kern-.125emX}}
\newsavebox{\policytablebox}
\newsavebox{\heavytailtablebox}
\newsavebox{\capacitytablebox}
\begin{document}

\title{Beyond Bot Labels: Retrospective Consequence-Grounded Ranking of Attributed Cache Harm in E-Commerce Logs
\thanks{Accepted to appear in the Proceedings of the 2026 IEEE International Conference on Big Data (IEEE BigData 2026). This is the author's accepted manuscript; the final version of record will be published by IEEE.}
\thanks{\textcopyright~2026 IEEE. Personal use of this material is permitted. Permission from IEEE must be obtained for all other uses, in any current or future media, including reprinting/republishing this material for advertising or promotional purposes, creating new collective works, for resale or redistribution to servers or lists, or reuse of any copyrighted component of this work in other works.}}

\author{\IEEEauthorblockN{Jisoo Kim}
\IEEEauthorblockA{\textit{Department of Data Science}\\
\textit{Sookmyung Women's University}\\
Seoul, Republic of Korea \\ sallysooo@sookmyung.ac.kr}
\and
\IEEEauthorblockN{Youngwoo Seo}
\IEEEauthorblockA{\textit{Department of Statistics}\\
\textit{Korea University}\\
Seoul, Republic of Korea \\ maple5795@korea.ac.kr}
\and
\IEEEauthorblockN{Seonghoon Jeong}
\IEEEauthorblockA{\textit{Division of Artificial Intelligence Engineering}\\
\textit{Sookmyung Women's University}\\
Seoul, Republic of Korea \\ seonghoon@sookmyung.ac.kr}}

\maketitle 
\bstctlcite{IEEEexample:BSTcontrol}

\begin{abstract}
Web caches reduce origin load and latency, but automated traffic can either warm them or evict objects shoppers reuse, lowering the cache hit ratio (CHR). Existing bot defenses generally infer automation from User-Agent (UA) strings or behavior rather than identify sessions that impose cross-session cache harm. We instead formulate the task as ranking sessions by eviction damage and propose consequence-grounded session ranking. A single chronological replay attributes each eviction-induced miss by a protected shopping session (a victim) to the session whose insertion last evicted the object. A behavioral model predicts the target from session behavior alone. Selected sessions bypass the cache but remain origin-served. In a controlled check of 539 non-victim sessions selected across harm levels, attribution order tracks victim-hit changes after removal (Spearman correlation 0.80). On a 12.7-million-request real trace, the top 1\% of sessions carry 48.8\% of attributed harm. Sessions without bot tags contribute 49.1\% of this tier's harm. At the 20\% request-volume target, the default conditional CHR difference reaches 93\% of the UA baseline, which selects 47 times as many sessions. Six training seeds show little variation. This comparison conditions on the victims that remain, so we also report victim bypass. Local replay and retraining on a second store reproduce harm concentration and a top tier dominated by sessions without bot tags. These results establish attributed cross-session harm as an outcome-based target for retrospective cache audits and selective intervention design.

\end{abstract}


\section{Introduction}
Shared web caches exploit request locality to reduce origin load and latency~\cite{arlitt1997trace,rfc9111}. These benefits matter in e-commerce, where latency is associated with lower conversion and sales~\cite{basalla2021latency,poggi2014responsetime}. Automated clients have disrupted cache locality for more than two decades~\cite{almeida2001robots,gao2006pollution}, and recent work estimates that they generate roughly half of web traffic~\cite{aicache}. Cloudflare reports that AI crawlers generated 20\% of its verified-bot traffic in 2025~\cite{cfradar}. Their cache effects differ. Crawlers revisiting popular pages may warm a cache. Scrapers sweeping cold Uniform Resource Identifiers (URIs) may instead evict objects reused by shoppers and lower the cache hit ratio (CHR).

Traffic-aware defenses commonly target either client identity or cache-locality heuristics. Bot-centric filters infer automation from User-Agent (UA) signatures or behavior. When labels are heuristic or synthetic, a classifier may learn the labeling procedure itself. Scrapers that use browser UAs can evade these filters, while identity-wide blocking can remove benign crawlers~\cite{tankumar,dorangokhale,hays,li2021goodbot,hemmatpour}. Cache-centric defenses instead use request repetition, a client's own hit rate, request clusters, or object popularity~\cite{deng2008pollution,manivel2003lrua,yao2020ddcpc,xie2012cacheshield}. The reviewed approaches do not directly attribute eviction-induced misses experienced by protected traffic. A client may retain a high own hit rate while its insertions evict objects that shoppers later request.

The retrospective task is to prioritize completed sessions for cache-bypass analysis, not to classify bots. Cache bypass sends selected requests to the origin without reading or writing the shared cache. It does not refuse them. Replaying once per candidate would evaluate its absence directly but is impractical at our scale. At the same request-volume target, a narrower session set reduces review burden and the surface for mistaken bypass. We therefore report session count and victim bypass alongside CHR (\cref{sec:evalproto,sec:payoff}).

Our consequence-grounded method ranks sessions by eviction-induced victim misses from one replay (\cref{alg:harm}). When a victim misses an evicted object, replay charges the session that last evicted it. We call this target consequence-grounded because it derives from misses experienced by protected sessions after an eviction rather than from identity, a session's own hit rate, or request volume. On the controlled stress-test trace, we check this ordering by removing 539 non-victim sessions selected across harm levels and replaying. It tracks the resulting victim-hit changes (Spearman $\rho=0.80$ in this sample, \cref{fig:validation}). A behavioral model predicts attributed harm without UA or role-tag inputs. We repeat attribution and training for each cache policy (\cref{sec:attrib,sec:predict}).

Ranking accuracy alone does not establish a useful bypass policy. We score completed sessions, replay each selected session as cache-bypassed from its first request, and report CHR for the victims that remain together with intervention breadth, victim bypass, and origin egress. The attribution check and injected-persona analyses use the controlled trace, whereas headline ranking results use real traffic (\cref{sec:evalproto,sec:policyharm}).

On our primary trace, EClog, attributed harm is concentrated, and many of the highest-harm sessions lack UA bot tags (\cref{tab:heavytail}). After local replay and training on a second store, harm remains concentrated and most top-tier sessions lack bot tags (\cref{tab:crossdomain}). At a common request-volume target, our default ranker reaches 93\% of the crawler-signature UA baseline's conditional CHR difference while the baseline selects 47 times as many sessions (\cref{tab:payoff}). The conditional difference stays positive across all tested policies and capacities, while better injected-persona detection changes conditional CHR only slightly (\cref{sec:policyharm,sec:decontam}).

The pipeline supports operational review by reporting intervention breadth and origin egress. Its outcome-based target also gives researchers a common measure for comparing defenses beyond automation accuracy.

This work makes the following three contributions:
\begin{itemize}\setlength{\itemsep}{0pt}\setlength{\parsep}{0pt}\setlength{\parskip}{0pt}
\item We provide cross-store evidence that attributed cache harm is concentrated and that UA tags leave a substantial share of the highest-harm tier uncovered (\cref{sec:heavytail,sec:crossdomain}).
\item We introduce a scalable ranking target by attributing eviction-induced victim misses in one replay. A controlled 539-session removal check supports its ordering, and a behavioral model predicts it without UA or role-tag inputs (\cref{sec:detection}).
\item We demonstrate a substantially narrower audit set. At EClog's default operating point, our ranker reaches 93\% of the UA baseline's conditional CHR difference while the baseline selects 47 times as many sessions. The evaluation jointly reports victim bypass and origin egress and tests policy, capacity, time, and store sensitivity (\cref{sec:results}).
\end{itemize}

The remainder is organized as follows. \Cref{sec:related,sec:data} review related work and describe data preprocessing. \Cref{sec:threat,sec:detection} present the stress-test design and method. \Cref{sec:results} reports results and discusses limitations. \Cref{sec:conclusion} concludes.

\section{Related Work}\label{sec:related}
Prior work addresses harmful automated traffic by identifying clients or protecting cache locality. We review both paths, then consequence attribution and the policies for evaluation.

\BfPara{Automated traffic and cache interference} Robots can increase cache misses by accessing broad object sets and disrupting locality~\cite{almeida2001robots}. Gao et al.~\cite{gao2006pollution} formalized locality disruption, which continually introduces unpopular objects, and false locality, which repeatedly accesses a small unpopular set. Controlled studies show analogous pressure in an in-network e-commerce cache~\cite{hemmatpour} and a Content Delivery Network (CDN) prototype under scan-like AI traffic~\cite{aicache}. These studies establish aggregate cache degradation but do not attribute victim misses to sessions in heterogeneous traffic.

\BfPara{Bot identification under weak labels} Web-robot detection has moved from UA and IP signatures toward navigation and behavioral features~\cite{tankumar,dorangokhale,xu2018,suchacka}. Recent methods use weakly supervised session models, graphs, and fingerprints~\cite{rovetta2020botrecognition,jagat2023webs4ae,stassopoulou2009probabilistic,dhamnani2020botcha,lagopoulos2020contentaware,semanticwebrobots2024,zhao2026graphbotdetection,wang2026fpagent}. They estimate identity or intent rather than cache consequence, and robot workloads are heterogeneous~\cite{doran2013comparison}. Identity evidence is also fragile. In a seven-month honeysite deployment, 86.2\% of bots claiming Chrome or Firefox had inconsistent TLS fingerprints, and most matched scripting libraries or command-line tools~\cite{li2021goodbot}. Browser-fingerprint labeling can likewise miss bots~\cite{iliou}, while \texttt{robots.txt} remains voluntary~\cite{liu2025somesite}. Because our logs contain cache keys but no page bodies, we treat identity labels as weak tags rather than cache-harm ground truth.

\BfPara{Traffic-aware cache protection} Prior defenses use per-client repetition and object lifetime~\cite{deng2008pollution}, recent own hits and misses~\cite{manivel2003lrua}, content-interest clusters~\cite{yao2020ddcpc}, or object popularity~\cite{xie2012cacheshield}. INCS instead learns a flow classifier from generated normal and bot traffic~\cite{hemmatpour}. Our adapted INCS-style baseline predicts attributed harm from timing and item popularity (\cref{sec:payoff}). \Cref{tab:relwork_baselines} summarizes these distinct analysis units and decision bases, none of which directly uses attributed cross-session harm.

\BfPara{Attributing shared-resource interference} The reviewed methods do not directly attribute a protected session's loss to another session. A related idea appears in processor-cache analysis: CASHT counts inter-core evictions to characterize shared last-level-cache contention and guide partitioning~\cite{gomes2022casht}. It studies concurrent compute workloads. We instead attribute later web-cache misses of a protected session population to earlier session insertions.

\begin{table}[t]
\caption{Comparison of traffic-aware cache defenses.}
\label{tab:relwork_baselines}
\centering
\small
\setlength{\tabcolsep}{3pt}
\begin{tabular}{@{}>{\raggedright\arraybackslash}p{1.00in}>{\raggedright\arraybackslash}p{0.78in}>{\raggedright\arraybackslash}p{1.48in}@{}}
\toprule
\textbf{Method} & \textbf{Analysis unit} & \textbf{Decision basis} \\
\midrule
Deng et al. (2008)~\cite{deng2008pollution} & client and object & repetition and lifetime \\
LRU-A (2003)~\cite{manivel2003lrua} & client & recent own hit/miss history \\
DDCPC (2020)~\cite{yao2020ddcpc} & content-interest cluster & content-interest clustering \\
CacheShield (2012)~\cite{xie2012cacheshield} & object & observed request count \\
INCS (2024)~\cite{hemmatpour} & flow & predicted synthetic bot class \\
\textbf{Ours} (2026) & session & predicted attributed harm \\
\bottomrule
\end{tabular}
\vspace{-1em}
\end{table}

\BfPara{General-purpose cache policies} Because attribution depends on replacement and admission, we also evaluate the Adaptive Replacement Cache (ARC)~\cite{megiddo2003arc}, SIEVE~\cite{sieve}, S3-FIFO~\cite{yang2023s3fifo}, and admission-filtered Least-Recently-Used (LRU). They respectively combine recency with frequency, use lazy FIFO promotion, filter one-hit objects through multiple FIFOs, or apply a simplified TinyLFU-style frequency rule~\cite{einziger2017tinylfu}. Among these, ARC, SIEVE, and S3-FIFO have been known to preserve human-facing hit ratio under synthetic AI scans even as aggregate hit ratio declines~\cite{aicache} (\cref{sec:evalproto,sec:policyharm}).

\BfPara{Relation to this work} Prior methods target automation or infer pollution from locality and demand. We instead attribute losses experienced by protected traffic, check the resulting session order by removal replay, and learn it from tag-independent behavior (\cref{sec:attrib,fig:validation}). The contribution is a consequence-grounded target and evaluation pipeline for selective bypass, not a bot classifier or replacement policy.

\section{Dataset and Preprocessing}\label{sec:data}

\begin{figure}[t]
\centering
\includegraphics[width=\columnwidth]{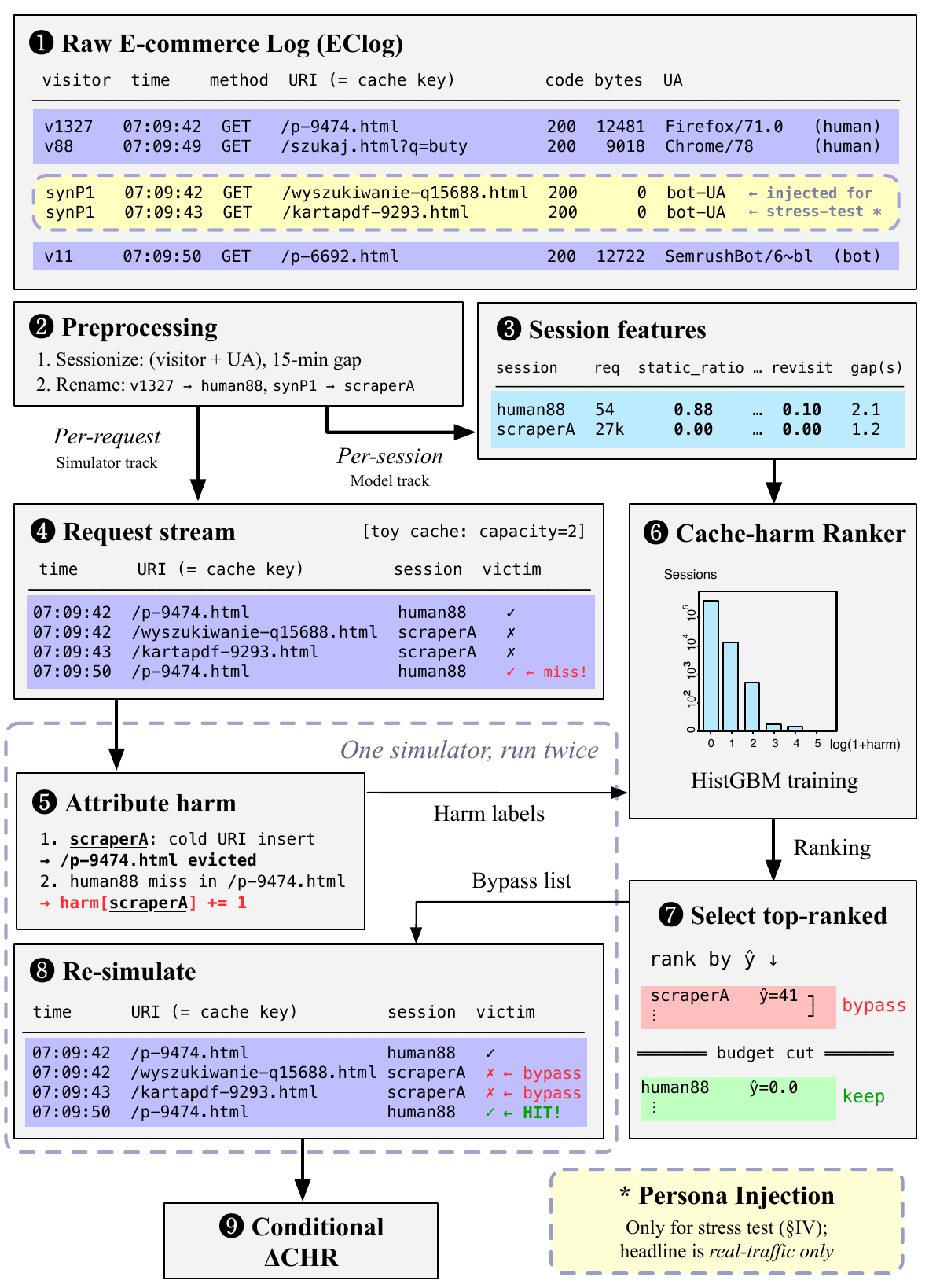}%
\caption{Pipeline from raw traffic (\stepicon{1}) through preprocessing, attribution, and ranking (\stepicon{2}--\stepicon{7}) to bypass replay and CHR evaluation (\stepicon{8}--\stepicon{9}). Injected traffic enters stress tests only.}
\label{fig:pipeline}
\vspace{-1em}
\end{figure}

Access logs expose request behavior and weak identity signals but not session-level cache consequences. We therefore construct two synchronized views: a chronological request stream for cache replay and a session table for attribution, ranking, and bypass. \Cref{fig:pipeline} shows how both views feed the remaining pipeline. This section defines the trace scope, constructs the views, and extracts the workload properties that motivate \cref{sec:threat,sec:detection}.

\subsection{Data Sources and Analytical Scope}
We use two public e-commerce traces for complementary purposes. EClog is the primary dataset for target construction and intervention comparisons~\cite{chodak}. It contains 35,157,691 raw requests recorded at one-second resolution over 183 days, from November 30, 2019 to May 31, 2020 (UTC), at a Polish e-commerce store. The documentation does not show whether upstream caches filtered this server log. Accordingly, we model the recorded stream rather than claim complete client demand. Zanbil is an independent five-day, 10.4-million-request trace from 2019~\cite{zaker2019zanbil}. After static-asset removal, its replay stream contains 1.94 million requests. Its adapter strips query strings from cache keys, whereas EClog retains them. We use Zanbil only for local reproduction and an identity audit (\cref{sec:crossdomain}). We neither pool the traces nor claim direct transfer.

\subsection{Request and Session Construction}
Cache state evolves per request, whereas harm attribution and intervention operate on sessions. We preserve request order while assigning every request to an analytical session.

\BfPara{Request processing} EClog encodes UTC time as .NET ticks and each client IP as a pseudonymous integer plus country code~\cite{chodak}. We convert time, separate the pseudonymous client identifier from its country code, and retain status codes 200, 206, 301, 302, 304, and 499. This status filter leaves 35,082,997 requests. Regular-expression rules map each URI to \texttt{product}, \texttt{browse}, \texttt{search}, \texttt{static\_asset}, \texttt{background}, or \texttt{other}. We exclude static assets from the replay stream but retain them as behavioral context. They form 63.9\% of the status-filtered stream. The resulting replay stream contains 12,665,505 requests across roughly 578,000 keys.

\BfPara{Session construction} We group consecutive requests by pseudonymous client-IP identifier and normalized UA, starting a new session after 15 minutes of inactivity. Prior EClog studies use raw UAs and a 30-minute gap~\cite{lim2023sessionclustering,lim2023clusterengage}. Relative to our construction, that gap produces 66\% fewer crawler-UA sessions but only 8\% fewer browser-UA sessions. These groups are analytical sessions, not verified users. The procedure yields 916,987 sessions, of which 635,082 contain replayed requests and enter harm attribution and ranking. \Cref{sec:policyharm} reports sensitivity to longer gaps.

\subsection{Weak Role Tags and Behavioral Features}
Weak tags and behavioral features serve different purposes. Tags specify candidate victims and the UA baseline. Features describe sessions without passing those tags to the model.

\BfPara{Weak role tags} We apply the priority rule \texttt{admin} $>$ \texttt{bot} $>$ \texttt{human} $>$ \texttt{unknown}. Administrative paths first define \texttt{admin}. Among the remaining sessions, crawler-signature UAs define \texttt{bot}, and browser UAs with shopping-intent requests define \texttt{human}. UA labeling is common but heuristic~\cite{tankumar,dorangokhale}. EClog's user-ID is empty for all but one request, and pseudonymous IP identifiers preclude reverse-DNS verification. The tags are not model inputs and do not determine our ranker's bypass list. They define candidate victims and thus which misses count toward attributed harm. A browser-UA scraper can satisfy the \texttt{human} rule, so victims receive an additional behavioral guard (\cref{sec:attrib,sec:decontam}).

\BfPara{Behavioral features} For prediction, we represent each session by 21 features. They include request volume and duration for scale. URI diversity, revisit ratio, top-URI share, and entropy capture locality. Inter-request-gap statistics and the share of zero-second gaps capture timing. URI-rarity statistics and hot-item ratio capture popularity. Static-asset, shopping-intent, and cacheable-content ratios capture composition. For URI $u$ and total request count $N$, URI rarity is $-\log_{10}(\operatorname{freq}(u)/N)$, using full-trace frequencies under the retrospective design (\cref{sec:limits}). UA strings, role tags, client identifiers, and country codes are not features. Because shopping intent helps define the \texttt{human} tag and one guard rule, the ranker excludes \texttt{intent\_ratio}, leaving 20 inputs (\cref{sec:attrib,sec:predict}).

\subsection{Workload Characteristics and Design Implications}\label{sec:workload}
The synchronized representations reveal three properties that explain why controlled stress and consequence-grounded ranking are both needed.

\BfPara{(1) A long cold tail limits rarity} Singleton keys account for 54.5\% of EClog and 71.1\% of Zanbil. Mean rarity rises from 3.3 to 5.6 between \texttt{human}- and \texttt{bot}-tagged EClog sessions and from 3.4 to 6.8 on Zanbil (\cref{tab:crossdomain}). Rarity separates these populations in aggregate but cannot distinguish a cache-harming session from an equally rare legitimate crawler. It can inform prediction but cannot serve as the consequence target.

\BfPara{(2) Observed crawler-UA sessions are low-intensity references} The median \texttt{bot}-tagged EClog session has seven requests, a 446-s mean gap, and no URI revisit. These sessions characterize background automation but provide no control over mechanism or intensity. \Cref{sec:threat} introduces higher-intensity personas for that purpose.

\BfPara{(3) Recorded bursts are not specific to automation} The median \texttt{human}-tagged session has a 0.86 static-asset ratio and a 0.9-s recorded mean gap, consistent with page-load bursts. EClog's one-second resolution cannot resolve timing within a second. Thus, timing alone is unsafe. The guard and ranker combine it with volume, locality, and content composition (\cref{sec:attrib,sec:predict}).

These properties motivate the controlled stress cases in \cref{sec:threat} and the consequence target in \cref{sec:detection}.

\section{Controlled Cache-Pollution Stress-Test Design}\label{sec:threat}
The traces provide realistic background traffic but do not reveal who or what generated the requests or let us control intensity. We therefore interleave three synthetic cache-polluting personas with EClog to isolate locality disruption and false locality~\cite{gao2006pollution,yao2020ddcpc}. Persona labels identify the injected sessions, whereas replay determines their cache harm. The controlled stream adds 15,000 sessions and 2,505,660 requests, none of which enter headline real-traffic results.

\BfPara{Persona design} P1 and P3 represent locality disruption at different scales and with different UA identities. P2 represents false locality:

\begin{itemize}
\item \textbf{P1 scan scraper} scans distinct cold URIs without static assets (median 73 requests, 0.5-s gaps, no revisits).
\item \textbf{P2 monitor bot} polls a small cold set to fabricate popularity (median 40 requests, 3.0-s gaps, 0.89 revisit ratio).
\item \textbf{P3 browser-spoofing scraper} scans distinct cold catalog and search URIs with a browser UA (median 380 requests, 1.1-s gaps, no revisits).
\end{itemize}

\BfPara{Injection procedure} To ground the stress cases in observed crawler demand while preserving their cold-key mechanisms, personas sample cacheable crawler-UA keys after excluding the hottest 30\%, with probability proportional to observed crawler request frequency. We generate 5,000 actors per persona, randomly vary volume, gaps, and revisit rate, place them at random times, and rebuild sessions. P1 and P2 use crawler-signature UAs and receive the weak \texttt{bot} tag. P3 receives the weak \texttt{human} tag because its browser UA and requests satisfy the tagging rule.

\BfPara{Calibration and scope} Against Zanbil crawlers, P1 and P3 volumes lie at the 94th and 98th percentiles, and P2-like high-revisit monitors occur naturally. Their request timing is deliberately aggressive: only 1\% of Zanbil crawlers are equally fast. Each persona is a stress case for a specific pollution mechanism, not a prevalence estimate. They support the policy and label-quality analyses (\cref{sec:policyharm,sec:decontam}). The next section derives harm targets for real sessions.

\section{Consequence-Grounded Harm Ranking}\label{sec:detection}
Controlled personas establish pollution mechanisms but do not identify harmful real sessions. We therefore attribute victim misses in one replay, check the order through sampled removals, predict it from tag-independent behavior, and compare rankings through full bypass replay. All stages are retrospective and use completed sessions.

\subsection{Protected Population and Eviction-Attributed Harm}\label{sec:attrib}
\BfPara{Protected population} A victim is a UA-human shopping session that passes a behavioral guard. We exclude a session from the victim set if it meets any of four rules. Rule A requires at least $300$ distinct URIs and a static-asset ratio below $0.05$. Rule B requires at least $200$ requests, a mean gap of at most $1$\,s, a revisit ratio below $0.2$, and a static ratio below $0.10$. Rule C requires a duration of at least $6$\,h, at least $500$ requests, a mean gap of at most $3$\,s, and a static ratio below $0.10$. Rule D requires at least $1000$ requests, a revisit ratio of at least $0.9$, and an intent ratio below $0.05$. The default set defines harm and the main CHR outcome. A stricter set requiring static ratio $\geq0.5$ and a looser set containing all UA-human sessions test only the sensitivity of the victim bypass rate.

Synthetic labels identify injected sessions rather than their cache effects and cannot serve as harm targets. We instead charge a victim miss to the session whose insertion most recently evicted the requested object. If session $A$ inserts $x$, evicts $y$, and a victim later misses $y$, the miss is charged to $A$, as formalized in \cref{alg:harm}. This ranking target requires one chronological replay rather than one replay per session.

The attribution is asymmetric. It records eviction damage without crediting cache warming and can overcharge sessions that do both. Victims are also ranked, and a victim can be charged for an object that its own earlier insertion evicted. The target can include same-session interference. These properties make attribution a ranking target rather than an estimate of harm prevented by removal. We check its ordering before learning from it and compare every bypass ranking through full replay. Harm is re-attributed for each cache policy. The first $10\%$ of each attribution stream is used only to warm cache state and does not contribute to harm counts.

\begin{algorithm}[t]
\caption{Single-pass session-level harm attribution under LRU.}
\label{alg:harm}
\begin{algorithmic}[1]
\REQUIRE time-ordered stream $(u_i,s_i,p_i)_{i=1}^{N}$: URI $u_i$, session $s_i$, and
    $p_i=1$ iff request $i$ belongs to a victim session (\cref{sec:attrib}); cache capacity $C$; warm-up length $w=\lfloor0.1N\rfloor$
\ENSURE per-session harm vector $\mathbf{h}$
\STATE $\textit{cache}\gets$ empty LRU order;\ \ $\textit{evictor}\gets\{\}$
\STATE $\mathbf{h}[s]\gets 0$ for every session $s$
\FOR{$i = 1,\dots,N$}
  \IF{$u_i \in \textit{cache}$}
    \STATE move $u_i$ to most-recently-used; remove $u_i$ from \textit{evictor}
    \STATE \textbf{continue} \COMMENT{cache hit: no miss to attribute}
  \ENDIF
  \IF{$p_i \;\wedge\; i > w \;\wedge\; u_i \in \textit{evictor}$}
    \STATE $\mathbf{h}[\,\textit{evictor}[u_i]\,] \mathrel{+}= 1$ \COMMENT{victim miss: charge the last evictor}
  \ENDIF
  \STATE insert $u_i$ as most-recently-used; remove $u_i$ from \textit{evictor}
  \IF{$|\textit{cache}| > C$}
    \STATE $x \gets$ least-recently-used item; evict $x$
    \STATE $\textit{evictor}[x] \gets s_i$ \COMMENT{$s_i$'s insertion pushed $x$ out}
  \ENDIF
\ENDFOR
\RETURN $\mathbf{h}$
\end{algorithmic}
\end{algorithm}

\subsection{Leave-One-Session-Out Replay Check}\label{sec:counterfactual}
Because attribution is not itself a removal effect, we test whether its ordering agrees with removal-based victim-hit changes. The check covers $539$ non-victim sessions selected across harm levels in the controlled stress-test trace. The sample contains $359$ positive-harm sessions, including the highest-harm cases and a random positive-harm draw, plus $180$ zero-harm controls. We remove each candidate in turn and replay. The change in counted victim hits is its leave-one-session-out value. Positive values mean that the removal replay records more victim hits. Because candidates are non-victims, their removal leaves all victim requests in the stream. The baseline and each removal replay apply the $10\%$ warm-up to their own request streams. The check covers this sample, not victim candidates or the full population.

\begin{figure}[t]
\centerline{\includegraphics[width=0.94\columnwidth]{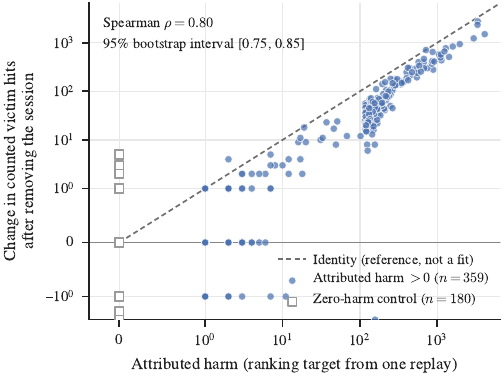}}
\caption{Removal check for 539 sampled non-victim sessions. Attributed harm tracks victim-hit change (Spearman $\rho=0.80$, 95\% bootstrap interval $[0.75,0.85]$). Axes are signed-log. The dashed line denotes identity.}
\label{fig:validation}
\vspace{-1em}
\end{figure}

\Cref{fig:validation} shows that, within this sample, attributed harm correlates with the leave-one-session-out victim-hit change across three orders of magnitude (Spearman $\rho=0.80$). Zero-harm controls have a median change of $0$, an interquartile range of $0$, and a mean of $-0.54$ hits. Correlation remains strong without zero-harm controls ($\rho=0.90$, $n=359$) or the highest-harm decile ($\rho=0.73$). The rankings share $17$ of their top $20$ sessions and $47$ of their top $50$. Because the sample was selected by attributed harm rather than at random, $\rho$ describes this sample rather than the full session population. We use attribution as a ranking target, not an estimate of harm prevented by removing a session.

\subsection{Predicting Harm from Behavior}\label{sec:predict}
Direct attribution requires replay-derived evictor records. We test how much of its order session behavior can recover before comparing both rankings under the same bypass simulation.

\BfPara{Objective} Scikit-learn's HistGradientBoostingRegressor~\cite{sklearn} 
maps the 20 behavioral inputs to $\log(1+\text{harm})$. Training uses all $635{,}082$ real replayed sessions, including human-tagged sessions and the $78\%$ with zero harm. Restricting training to UA bots would let the weak tag define learnable harm. The excluded \texttt{intent\_ratio} does not change the ranking, and synthetic personas do not enter this model.

\BfPara{Protocol} The regressor uses 400 iterations, learning rate $0.06$, and $L_2=1.0$. Five-fold out-of-fold prediction means that each session is scored by a model not trained on that session. The default split is by session. Cache state and full-trace statistics are shared across folds, so the folds are not fully independent. An alternative split keeps each pseudonymous-client--UA actor in one fold (\cref{sec:policyharm}).

\BfPara{Prediction quality} These scores reach Spearman $\rho=0.49$ with attributed harm even though $78\%$ of sessions share the same zero-harm value. The top $5\%$ by predicted score capture $63\%$ of harm, versus $69\%$ for the attributed-harm reference.

\BfPara{Feature contribution} Permutation importance ranks static-asset ratio first, followed by mean URI rarity and top-URI share. Request count ranks fifth, consistent with the nonpositive bypass result for request-volume ranking (\cref{sec:payoff}).

\subsection{Retrospective Bypass Simulation and Evaluation Protocol}\label{sec:evalproto}
We compare the learned ranker, direct attribution, and baselines under the same simulated action, reporting cache outcome with intervention breadth, victim bypass, and origin cost. Bypassed-request volume is selected-session requests divided by all replay requests.

\BfPara{Simulated bypass} Selected sessions remain origin-served but neither read nor write the shared cache. A false positive incurs an uncached response rather than denial of service.

\BfPara{Metrics} We report CHR differences in percentage points (pp). Let $V$ denote all original victims and $V_m$ those not bypassed by method $m$. Conditional $\Delta$CHR is post-bypass CHR for $V_m$ minus no-bypass CHR for $V$. It decomposes as 
\begin{equation}
[\mathrm{CHR}_m(V_m)-\mathrm{CHR}_0(V_m)]+[\mathrm{CHR}_0(V_m)-\mathrm{CHR}_0(V)].
\label{eq:chr-decomp}
\end{equation}

The first term captures the cache-state difference for retained victims, and the second captures population composition after victim bypass. We therefore interpret conditional $\Delta$CHR as their joint operating-point outcome and report victim bypass beside it. 
At the default operating point, this decomposition attributes $+0.74$\,pp of the reported $+1.04$\,pp gain to genuine cache-state improvement for retained victims and the remaining $+0.31$\,pp to the shift in which victims are counted after bypass.
Victim-origin egress counts response bytes on misses from $V_m$ over the same per-replay window as CHR. Total origin egress counts the full replay and includes every bypassed response. Victim bypass rate is $|V\setminus V_m|/|V|$ (\cref{sec:payoff}).

\BfPara{Replay} We replay requests chronologically from an empty cache, and each miss inserts its URI key~\cite{arlitt1997trace}. The first $10\%$ warms cache state but is excluded from harm and CHR. Bypass and removal replays recompute this boundary on the retained stream, whereas attribution uses the full-stream boundary. The fraction is common across policies.

\BfPara{Cache configuration} Capacity defaults to $1\%$ of approximately 578,000 replay-stream URI keys, or about 5,780 unit-object entries. A smaller P1/P2 sweep selected this point because injected personas reduced victim CHR by $1.14$\,pp, versus $0.65$\,pp at $5\%$. The full stress test reaches $4.26$\,pp (\cref{tab:policy}). We vary capacity and keep each replay setting common across methods. Admission LRU compares periodically halved exact counts for a candidate and the LRU eviction candidate~\cite{einziger2017tinylfu}. ARC uses integer adaptation, and our S3-FIFO uses small, main, and ghost FIFOs with one capped frequency counter.

\BfPara{Cacheability} The default is a simplified URI-key application cache without time-to-live (TTL), not a standards-compliant HTTP cache. Applications may cache beyond conventional method and status rules~\cite{mertz2018appcaching}, but POST and 304 have specific semantics~\cite{rfc9110,rfc9111}. We therefore also retain the original sessions and features while rebuilding a GET/200-only stream, $1\%$ capacity, harm attribution, and local model.

\BfPara{Budgets and baselines} The main evaluator includes the first ranked session that reaches the volume target, so session granularity can slightly exceed it. Temporal and Zanbil checks use the largest prefix below their caps and report realized volume. Session count and victim bypass remain separate costs. The UA baseline orders regex-tagged \texttt{bot} sessions by volume and applies the same bypass action. LRU-A uses its threshold-selected point. A seeded random ordering supplies the main floor, and the temporal check averages five orderings. The attributed-harm reference directly follows \cref{alg:harm}. It is not an oracle because attribution omits warming credit and multi-session removal interactions. Its gap to our ranker measures prediction cost.

\section{Experiments and Results}\label{sec:results}
The evaluation addresses four research questions (RQs) that provide the evidence needed to justify an actionable ranking. The first two RQs test whether attributed harm provides a selective target that recurs beyond the primary trace. The last two test whether learning that target narrows intervention under varied settings and whether better identity detection produces the same cache outcome.

\RQBox{1}{Is attributed harm concentrated, and how much of the high-harm tier do UA tags miss?}
\RQBox{2}{Do harm concentration and incomplete UA-tag coverage recur on a second store?}
\RQBox{3}{Does the learned ranking remain selective across tested settings?}
\RQBox{4}{Does better injected-persona detection improve cache outcomes?}

Unless stated otherwise, results use real EClog traffic, LRU, $1\%$ capacity, and a $20\%$ bypassed-request-volume target. We mark injected, alternative, and non-volume-matched results.

\subsection{Harm Concentration and UA-Tag Coverage}
\label{sec:heavytail}
We test whether concentration supports selective intervention and whether UA tags cover EClog's high-harm tier.

\begin{figure}[t]
\centerline{\includegraphics[width=0.84\columnwidth]{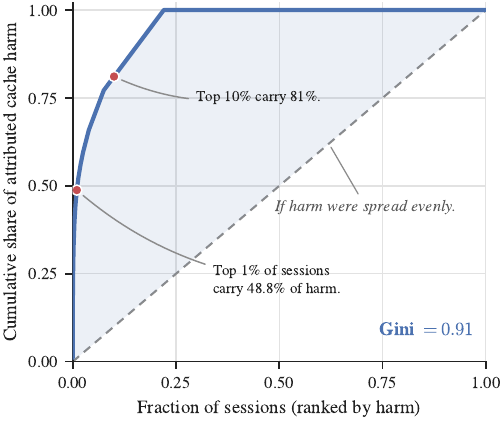}}
\caption{Attributed-harm concentration on EClog under $1\%$-capacity LRU. The top $1\%$ and $10\%$ carry $48.8\%$ and $81\%$ of harm (Gini $0.91$). The diagonal denotes equality.}
\label{fig:lorenz}
\vspace{-1em}
\end{figure}

\BfPara{Harm concentration} Among the 635,082 replay-participating EClog sessions, the top $1\%$ and $10\%$ account for $48.8\%$ and $81\%$ of attributed harm, respectively, with Gini $0.91$ indicating strong concentration (\cref{fig:lorenz}). This concentration motivates selective auditing but does not show that UA labels find the responsible sessions.

\BfPara{UA-tag coverage} The \texttt{bot} tag covers only about half of top-tier harm (\cref{tab:heavytail}). Among the uncovered roles, $75$ browser-UA sessions span hundreds to tens of thousands of distinct URIs over multi-hour activity and contribute $11\%$ of all attributed harm. This subset is a descriptive inspection, not a predefined scraper rule or verified identity class. The tag-level totals supply the reproducible UA comparison.

\RQBox{1}{\textbf{Answer.} Yes. The top $1\%$ carries $48.8\%$ of harm, with $49.1\%$ of that tier from sessions without bot tags.}

\begin{table}[t]
\caption{Role tags in EClog's top $1\%$ by attributed harm ($n=6{,}351$).}
\label{tab:heavytail}
\centering
\setlength{\tabcolsep}{5pt}
\sbox{\heavytailtablebox}{\begin{tabular}{l>{\centering\arraybackslash}p{1.3cm}>{\centering\arraybackslash}p{1.3cm}}
\toprule
 & \multicolumn{2}{c}{\textbf{Share of the top $1\%$}} \\
\cmidrule(l){2-3}
\textbf{Role} & \textbf{Sessions} & \textbf{Harm} \\
\midrule
\texttt{bot} (caught by UA label)   & 59.0\% & 50.8\% \\
\texttt{human} (not labeled a bot)  & 33.9\% & 37.6\% \\
\texttt{admin}                      & \phantom{0}5.2\% & 10.7\% \\
\texttt{unknown}                    & \phantom{0}1.8\% & \phantom{0}0.8\% \\
\midrule
\textit{Not labeled as bots} (role $\neq$ \texttt{bot}) & 41.0\% & 49.1\% \\
\quad browser-UA high-breadth subset ($n=75$) & \phantom{0}1.2\% & 23.3\% \\
\bottomrule
\end{tabular}}%
\usebox{\heavytailtablebox}
\vspace{-1.5em}
\end{table}

\subsection{Cross-Store Reproduction on a Second E-Commerce Log}
\label{sec:crossdomain}
A single-store pattern could reflect EClog's workload or tagging rules. We therefore apply the same session construction to Zanbil~\cite{zaker2019zanbil} and rerun victim selection, attribution, and training locally. No EClog model is transferred.

\BfPara{Reproduced phenomena} Zanbil has a larger singleton share, wider UA-tagged rarity gap, and more concentrated harm than EClog (\cref{tab:crossdomain}). Most of its top tier lacks the \texttt{bot} tag, and a descriptively similar browser-UA high-breadth subset again carries a notable all-harm share. The phenomena recur, but their magnitudes are workload-dependent.

\BfPara{Local bypass simulation} We compare EClog's $20\%$ point with Zanbil's largest prefixes below a $21\%$ cap (\cref{tab:crossdomain}). The points are not volume-matched, and victim bypass stays below $0.3\%$ in both stores. On Zanbil, our ranker selects 95 sessions versus 2 for UA-based bypass because one crawler supplies one-eighth of the stream and prevents a closer match. The result supports local refitting without bot-tag targets or inputs, not direct transfer or stable session counts across stores.

\BfPara{Identity audit} Because this comparison relies on weak role tags, Zanbil also permits a limited audit of their identity evidence. Among 508 crawler-UA clients with successful forward-confirmed reverse DNS~\cite{googleCrawlerVerification}, four are inconsistent with their claimed operator. Separately, $0.4\%$ of browser-UA clients map to hosting or datacenter networks. Unresolved clients are excluded, and the DNS data were collected after capture. This selected audit therefore does not estimate overall identity-label precision or coverage. Hosting or datacenter mapping does not prove automation.

\RQBox{2}{\textbf{Answer.} Yes. Zanbil reproduces concentrated harm and a top tier dominated by sessions without bot tags, with workload-dependent magnitude and breadth.}

\begin{table}[t]
\caption{Cross-store results and store-local bypass simulations.}
\label{tab:crossdomain}
\centering
\setlength{\tabcolsep}{3pt}
\begin{tabular}{lcc}
\toprule
 & \textbf{EClog} & \textbf{Zanbil} \\
\midrule
\multicolumn{3}{l}{\textit{Observed phenomena}} \\
Singleton-URI share (cold tail)          & $54.5\%$ & $71.1\%$ \\
Mean URI rarity, \texttt{human}$\to$\texttt{bot} & $3.3\!\to\!5.6$ & $3.4\!\to\!6.8$ \\
Gini of per-session attributed harm      & $0.91$ & $0.925$ \\
Top-$1\%$ attributed-harm share          & $48.8\%$ & $67\%$ \\
Top-$1\%$ sessions without \texttt{bot} tag & $41\%$ & $73\%$ \\
All-harm share, high-breadth browser UAs  & $11\%$ & $8.3\%$ \\
\midrule
\multicolumn{3}{l}{\textit{Store-local conditional $\Delta$CHR (pp)}} \\
\quad Attributed-harm reference & $+1.20$ & $+1.04$ \\
\quad \textbf{Harm ranking (ours)}  & $\mathbf{+1.04}$ & $\mathbf{+1.43}$ \\
\quad UA-based bypass          & $+1.12$ & $+1.04$ \\
\quad Volume bypassed, ours / UA & $20\%$ / $20\%$ & $13.8\%$ / $17.3\%$ \\
\quad Sessions bypassed, ours / UA & $3.6$k / $167$k & $95$ / $2$ \\
\bottomrule
\end{tabular}
\end{table}

\subsection{Bypass Outcomes at a Common Request-Volume Target}\label{sec:payoff}
\begin{figure*}[t]
\centering
\subfloat[Conditional $\Delta$CHR versus intervention breadth.\label{fig:operating-a}]{\includegraphics[width=0.48\textwidth]{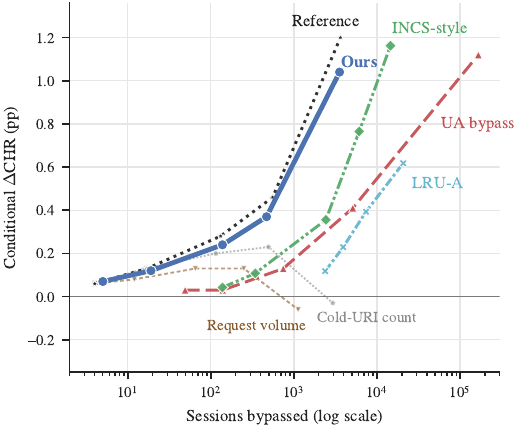}}
\hfill
\subfloat[Conditional $\Delta$CHR versus victim bypass rate.\label{fig:operating-b}]{\includegraphics[width=0.48\textwidth]{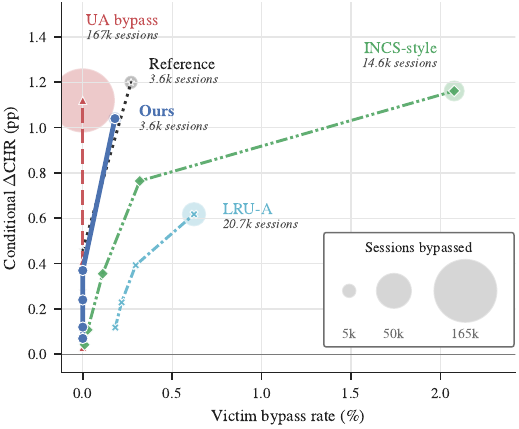}}
\caption{EClog conditional $\Delta$CHR operating curves under LRU with a $1\%$ cache. Breadth spans $1$--$20\%$ bypassed volume. Bubble area denotes sessions. UA-based bypass has zero victim bypass by construction.}
\label{fig:operating}
\vspace{-1em}
\end{figure*}

The cross-store check tests recurrence but does not provide a volume-matched method comparison. We return to real EClog traffic and use retrospective full-session replay under LRU with a $1\%$ cache. In \cref{tab:payoff,fig:operating}, budgeted rankings share a $20\%$ volume target. LRU-A~\cite{manivel2003lrua} uses threshold-selected points. Injected personas are excluded.

\BfPara{Conditional $\Delta$CHR and breadth} Our ranker reaches 93\% of the UA baseline's conditional $\Delta$CHR while UA-based bypass selects 47 times as many sessions (\cref{tab:payoff}). Six training seeds yield $+1.02\pm0.02$\,pp, showing a stable operating point within this trace. The $0.16$-pp gap to the attributed-harm reference quantifies the cost of predicting rather than directly using the target (\cref{sec:evalproto}).

\BfPara{Heuristics} Volume, cold-URI (each session's distinct-URI total), and random rankings yield nonpositive conditional $\Delta$CHR. Mean rarity reaches $+4.27$\,pp but bypasses $28.4\%$ of victims, whose baseline CHR is much lower than that of remaining victims (EClog $0.09$ vs.\ $0.85$ and Zanbil $0.08$ vs.\ $0.62$). This is consistent with survivor selection rather than targeted mitigation.

\BfPara{Adapted baselines} The INCS-style ranker~\cite{hemmatpour} predicts our target from \texttt{mean\_gap\_sec} and \texttt{mean\_uri\_rarity}. LRU-A uses its threshold-selected point (\cref{tab:payoff}). Both select broader sets, and neither adaptation reproduces the original system end to end.

\BfPara{Victim overlap} A narrow selected set is useful only if it also limits overlap with the protected population. Under the default guard, the victim bypass rate is $0.18\%$, and these bypassed victims issue $5.9\%$ of victim requests. UA-based bypass has a zero victim bypass rate by construction because the \texttt{bot} and \texttt{human} tags are disjoint.

\BfPara{Origin load} Session breadth and victim bypass do not capture origin cost. Although conditional victim-origin egress falls (\cref{tab:payoff}), total egress from a $66.06$\,GB baseline rises $12.6\%$ for ours and $11.5\%$ for UA-based bypass. No-admit limits our increase to $1.8\%$ but yields $-0.65$\,pp conditional $\Delta$CHR because selected-session hits still refresh recency. Refusal avoids this load but turns false positives into denial of service.

\begin{table}[t]
\caption{EClog bypass outcomes at $20\%$ volume (LRU-A: threshold-selected $8.3\%$). Victim-egress reduction compares post-bypass $V_m$ origin bytes with no-bypass $V$.}
\label{tab:payoff}
\centering
\setlength{\tabcolsep}{3pt}
\begin{tabular}{lr>{\centering\arraybackslash}p{1.15cm}>{\centering\arraybackslash}p{1.15cm}>{\centering\arraybackslash}p{1.15cm}}
\toprule
\textbf{Ranking} & \textbf{Sessions} & \textbf{Victim} & \textbf{Cond.} & \textbf{Victim} \\
 & \textbf{bypassed} & \textbf{bypass} & \textbf{$\Delta$CHR (pp)} & \textbf{egress red.} \\
\midrule
Attributed-harm reference & $3{,}637$   & $0.27\%$ & $+1.20$ & $5.8\%$ \\
\textbf{Harm ranking (ours)} & $3{,}551$ & $0.18\%$ & $+1.04$ & $5.2\%$ \\
UA-based bypass & $166{,}622$ & $0.00\%$ & $+1.12$ & $3.7\%$ \\
LRU-A~\cite{manivel2003lrua} & $20{,}737$ & $0.62\%$ & $+0.62$ & $2.0\%$ \\
INCS-style~\cite{hemmatpour} & $14{,}559$ & $2.08\%$ & $+1.16$ & $7.8\%$ \\
\midrule
Cold-URI count       & $2{,}944$   & $0.80\%$ & $-0.03$ & $5.2\%$ \\
Request volume       & $1{,}128$   & $0.09\%$ & $-0.06$ & $2.1\%$ \\
Mean URI rarity      & $277{,}798$ & $28.4\%$ & $+4.27$ & $20.4\%$ \\
Random baseline         & $130{,}170$ & $20.5\%$ & $-0.08$ & $32.7\%$ \\
\bottomrule
\end{tabular}
\end{table}

\begin{table}[t]
\caption{Capacity sensitivity at the $20\%$ volume target.}
\label{tab:capacity}
\centering
\setlength{\tabcolsep}{3pt}
\sbox{\capacitytablebox}{\begin{tabular}{l>{\centering\arraybackslash}p{1.25cm}>{\centering\arraybackslash}p{0.85cm}>{\centering\arraybackslash}p{0.95cm}>{\centering\arraybackslash}p{1.55cm}}
\toprule
 & & \multicolumn{2}{c}{\textbf{Cond. $\Delta$CHR (pp)}} & \\
\cmidrule(lr){3-4}
\textbf{Capacity} & \textbf{No-bypass CHR} & \textbf{Ours} & \textbf{UA-based} & \textbf{Sessions (Ours\,/\,UA)} \\
\midrule
$0.5\%$      & $0.850$ & $+1.06$ & $+1.58$ & $3.5$k\,/\,$167$k \\
$1\%$ (main) & $0.874$ & $+1.04$ & $+1.12$ & $3.6$k\,/\,$167$k \\
$5\%$        & $0.920$ & $+0.59$ & $-0.27$ & $3.4$k\,/\,$167$k \\
\bottomrule
\end{tabular}}%
\usebox{\capacitytablebox}
\vspace{-1em}
\end{table}

\subsection{Sensitivity to Session, Model, and Cache Settings}
\label{sec:policyharm}
We group sensitivity into cache policy, analytical construction (model split, session gap, and victim boundary), temporal ordering, and replay assumptions (capacity, byte weighting, and cacheability). In \cref{tab:policy}, loss uses the injected stream and conditional $\Delta$CHR uses real EClog traffic. Policy tests use $1\%$ capacity and a $20\%$ volume target, except threshold-selected LRU-A. We re-attribute harm and retrain for every policy.

\BfPara{Policy effects} The personas leave $3.48$--$4.26$\,pp of clean-minus-attacked victim-CHR loss. SIEVE has the smallest loss, but none eliminates it (\cref{tab:policy})~\cite{sieve,aicache}. Policy-local rankers remain positive while UA-based bypass selects 45--63 times as many sessions, although ours leads only under Least-Frequently-Used (LFU) eviction. Under SIEVE, ARC, S3-FIFO, and admission LRU, even direct attributed-harm ordering trails UA-based bypass, exposing a surrogate limitation rather than prediction error. The mechanisms also differ: P2's repetition harms LFU, whereas admission LRU produces few eviction charges despite CHR loss. Adapted LRU-A~\cite{manivel2003lrua} selects $7.0$--$10.7\%$ volume and trails both rankers under every policy, reaching $-0.77$\,pp when its no-admit action overlaps admission LRU.

\BfPara{Other defenses} No non-trivial CacheShield-style configuration~\cite{xie2012cacheshield} exceeds LRU in our clean-traffic sweep, which shows cost here rather than ineffectiveness under attack. A Deng-style session score~\cite{deng2008pollution} yields $-0.25$\,pp because it selects high-repeat, high-hit sessions, versus $-0.08$\,pp for random.

\BfPara{Model-split and session-gap sensitivity} Keeping each client--UA actor in one fold yields $+1.16$\,pp, versus $+1.04$\,pp for session folds. At 30- and 60-minute gaps, ours yields $+0.93$ and $+1.01$\,pp with 2.8k and 0.8k sessions. UA-based bypass yields $+1.15$\,pp with 72.4k and 33.6k. Victim bypass is at most $0.35\%$.

\BfPara{Victim-boundary sensitivity} Stricter and looser boundaries keep the victim bypass rate below $0.25\%$, so the low session-level overlap persists across the tested definitions. Because the ranker and guard share inputs, including static-asset ratio, we interpret this rate as guard-conditioned operational scope rather than identity accuracy. This check varies victim bypass only, not CHR (\cref{sec:limits}).

\BfPara{Temporal sensitivity} Training on the first $80\%$ and testing on the final month yields $+1.00$\,pp versus $+0.84$\,pp for UA-based bypass, with 708 versus 43k sessions. A static high-harm list improves the difference only with much higher victim bypass, favoring periodic rescoring. Because URI rarity still uses the full trace, this is a partial temporal test. A prospective pipeline remains future work (\cref{sec:limits}).

\BfPara{Capacity} Ours remains positive across $0.5$--$5\%$ capacity with victim bypass at most $0.35\%$ (\cref{tab:capacity}). UA-based bypass leads at smaller capacities but becomes negative at $5\%$, consistent with crawler requests warming the no-bypass cache.

\BfPara{Byte weighting} Count- and byte-based targets can induce different orders. With each URI's maximum response size and capacity at $1\%$ of catalog bytes, direct object-count ordering yields $+1.02$\,pp versus $+1.42$\,pp for byte-weighted attribution. This sensitivity supports rebuilding the target under local object costs before training.

\BfPara{Cacheability} Because the default includes $43\%$ POST traffic and 304/499 responses, we repeat the analysis on GET/200 only. Harm stays concentrated (Gini $0.88$, top-$1\%$ share $45.8\%$), but session order changes (Spearman $\rho=0.41$). After local retraining, ours yields $+1.68$\,pp with 1.9k sessions, versus $+2.03$\,pp with 32.3k for UA-based bypass. The breadth ratio falls from $47\times$ to $17\times$, and the relative conditional $\Delta$CHR from $93\%$ to $83\%$. Thus selectivity persists, but the target and ratio depend on cacheability.

\RQBox{3}{\textbf{Answer.} Yes. Ours remains positive across tested policies and capacities. At the default point, it reaches 93\% of UA-based conditional $\Delta$CHR while UA selects 47 times as many sessions.}

\begin{table}[t]
\caption{Injected victim-CHR loss (clean minus attacked) and real-traffic conditional $\Delta$CHR by policy ($1\%$ capacity).}
\label{tab:policy}
\centering
\setlength{\tabcolsep}{2pt}
\sbox{\policytablebox}{\begin{tabular}{l>{\centering\arraybackslash}p{0.80cm}>{\centering\arraybackslash}p{0.90cm}>{\centering\arraybackslash}p{0.90cm}>{\centering\arraybackslash}p{0.90cm}>{\centering\arraybackslash}p{1.35cm}}
\toprule
 & & \multicolumn{3}{c}{\textbf{Cond. $\Delta$CHR (pp)}} & \\
\cmidrule(lr){3-5}
\textbf{Policy} & \textbf{Loss (pp)} & \textbf{Ours} & \textbf{UA} & \textbf{LRU-A} & \textbf{UA/Ours sessions} \\
\midrule
LRU            & $4.26$ & $+1.04$ & $+1.12$ & $+0.62$ & $47\times$ \\
LFU            & $4.24$ & $+2.01$ & $+1.62$ & $+1.01$ & $61\times$ \\
SIEVE          & $3.48$ & $+0.76$ & $+1.11$ & $+0.22$ & $46\times$ \\
Admission LRU  & $3.81$ & $+1.29$ & $+1.49$ & $-0.77$ & $63\times$ \\
ARC            & $3.84$ & $+1.06$ & $+1.35$ & $+0.37$ & $45\times$ \\
S3-FIFO        & $3.54$ & $+0.77$ & $+0.98$ & $+0.23$ & $46\times$ \\
\bottomrule
\end{tabular}}
\usebox{\policytablebox}
\vspace{-1em}
\end{table}

\subsection{Auxiliary Audit of UA-Label Contamination}\label{sec:decontam}
Identity accuracy need not predict cache consequence. We test this distinction by cleaning a separate detector's normal set, then comparing injected-persona detection with real-traffic bypass.

\BfPara{Detector and setup} The audit uses a one-class long short-term memory (LSTM) autoencoder~\cite{malhotra} over five-request windows (hidden $64$, latent $16$, Adam $10^{-3}$). Inputs are URI type, log gap, rarity, status, and followed-static count. Trained on UA-human sessions, it fixes the window-error threshold at a $5\%$ held-out false-positive rate and ranks sessions by anomalous-window fraction. Short sessions are zero-padded. The detector is independent of the harm ranker's training target and bypass list. The contaminated set contains all UA-human sessions. The decontaminated set removes sessions failing the victim guard. UA strings and role tags select training windows but are not features, so browser-UA P3 enters the contaminated set. A real UA-human session with $39{,}000$ distinct URIs over $16$ hours illustrates extreme behavior, not proven automation.

\BfPara{Detection and bypass effects} Across five seeds, filtering raises LSTM area under the receiver operating characteristic curve (ROC-AUC) from $0.77\pm0.04$ to $0.92\pm0.02$. This measures injected-persona separation, not identity accuracy on real traffic. On real EClog at the $20\%$ target, mean conditional $\Delta$CHR changes from $+0.98\pm0.03$ to $+1.00\pm0.04$\,pp (\cref{tab:decontam}). The change is small relative to seed variation, but does not establish equivalence. The detector still selects $153$--$154$k sessions and $3.15$--$3.28\%$ of victims. Across six seeds, the harm ranker reaches $+1.02\pm0.02$\,pp with $3.5$k sessions and a $0.18\%$ rate.

\RQBox{4}{\textbf{Answer.} No proportional cache gain appears. Decontamination raises injected-persona ROC-AUC by $0.15$ but changes mean conditional $\Delta$CHR by only $0.02$\,pp and still selects broadly.}

\begin{table}[t]
\caption{Five-seed detector and six-seed harm-ranking outcomes (mean$\pm$standard deviation).}
\label{tab:decontam}
\centering
\setlength{\tabcolsep}{3.5pt}
\begin{tabular}{lcrrr}
\toprule
\textbf{Method} & \textbf{ROC-AUC} & \shortstack{\textbf{Cond. $\Delta$CHR}\\\textbf{(pp)}} & \shortstack{\textbf{Sessions}\\\textbf{bypassed}} & \shortstack{\textbf{Victim}\\\textbf{bypass}} \\
\midrule
Contaminated LSTM & $0.77 \pm .04$ & $+0.98 \pm 0.03$ & $153$k & $3.28\%$ \\
Decontaminated LSTM & $\mathbf{0.92 \pm .02}$ & $+1.00 \pm 0.04$ & $154$k & $3.15\%$ \\
\midrule
\textbf{Harm ranking (ours)}    & --- & $\mathbf{+1.02 \pm 0.02}$ & $\mathbf{3.5}$\textbf{k} & $\mathbf{0.18\%}$ \\
\bottomrule
\end{tabular}
\end{table}

\subsection{Discussion and Limitations}\label{sec:limits}
The results support consequence-grounded ranking as a selective retrospective audit, with three limits.

\BfPara{Retrospective scope} Conditional $\Delta$CHR combines cache-state change with survivor selection. It is an operating-point outcome, not a causal recovery estimate. Sessions are scored after completion, and labels, cache state, and URI rarity use the full trace. A preliminary running-threshold variant of the detector underperformed offline selection on injected traffic and still used corpus-global rarity. Prospective use requires past-only features and live validation.

\BfPara{Attribution and action} Attributed harm is cache- and policy-dependent, omits warming credit, and may include same-session interference. It is not a removal estimate, and its removal check covers sampled non-victims. Replay omits expiration and cache tiers. We do not test latency, monetary cost, or adaptive evasion. Bypass raises origin egress, requiring local retraining and origin-aware budgets.

\BfPara{Population and transfer} Victims and UA-derived roles are heuristic. Alternative boundaries test victim bypass, not CHR or identity. The 75-session subset is descriptive, sessions are not people, and P3 is not a prevalence estimate~\cite{aicache,cfradar}. EClog documents no secondary key for one-second timestamp ties. Replay sorts only by timestamp. Identifiers support aggregate analysis only.

\section{Conclusion}\label{sec:conclusion}
Bot labels target automation, not cross-session cache consequence. We introduced consequence-grounded ranking, which turns eviction-induced victim misses from one replay into a session target and predicts it from tag-independent behavior. In a controlled sample of 539 non-victim sessions, the order tracks victim-hit changes under removal (Spearman $\rho=0.80$).

Across two stores, harm is concentrated and many top-harm sessions lack bot tags. At EClog's default target, our ranker reaches 93\% of the UA baseline's conditional CHR difference, while the baseline selects 47 times as many sessions. Ours bypasses $0.18\%$ of victims. The difference stays positive across tested policies and capacities, but its setting-dependent magnitude requires local attribution and replay.

These results give researchers a consequence-based target for comparing cache defenses beyond identity accuracy. For operations, the pipeline prioritizes a narrow audit set while exposing victim overlap and origin cost. Consequence-grounded ranking can therefore complement bot management in retrospective reviews and guide selective intervention design. Prospective scoring, local retraining, and origin-aware budgets are the next steps toward live use.

\let\oldthebibliography\thebibliography
\renewcommand\thebibliography[1]{\oldthebibliography{#1}\footnotesize\setlength{\itemsep}{0pt}\setlength{\parskip}{0pt}}
\bibliographystyle{IEEEtran}
\bibliography{references}

\end{document}